\documentclass[
aps,
pra,
notitlepage,
citeautoscript, superscriptaddress,
reprint
]{revtex4-2}
\usepackage[dvipsnames,usenames]{xcolor}
\usepackage{graphicx}
\usepackage{graphicx}
\usepackage{dcolumn}
\usepackage{bm}
\usepackage{mathrsfs,natbib}
\usepackage[title]{appendix}
\usepackage{graphicx,epsf,amssymb,amsbsy,amsfonts,amssymb,amsmath}
\usepackage[colorlinks=true]{hyperref}
\hypersetup{
    unicode=false,          
    pdftoolbar=true,        
    pdfmenubar=true,        
    pdffitwindow=false,     
    pdfstartview={FitH},    
    pdftitle={},    
    pdfauthor={},     
    pdfsubject={},   
    pdfcreator={},   
    pdfproducer={}, 
    pdfkeywords={} {} {}, 
    pdfnewwindow=true,      
    colorlinks=true,       
    linkcolor=magenta, 
    citecolor=blue,        
    filecolor=magenta,      
    urlcolor=blue           
}
\newcommand{\be}{\begin{equation}}
\newcommand{\ee}{\end{equation}}
\newcommand\beq{\begin{eqnarray}}
\newcommand\eeq{\end{eqnarray}} 

\begin{document}

\title{Unpaired Weyl fermion on an axion string in a finite lattice}
\author{Jonathan D. Kroth}%
\email{jdkroth@iastate.edu}
\affiliation{Department of Physics and Astronomy, Iowa State University, Ames, Iowa 50011, USA}%
\author{Srimoyee Sen}%
\email{srimoyee08@gmail.com}
\affiliation{Department of Physics and Astronomy, Iowa State University, Ames, Iowa 50011, USA}%

\date{\today}

\begin{abstract}
Domain wall fermions use a $2n$-dimensional spacetime defect embedded in $(2n+1)$-dimensional spacetime to realize massless lattice Dirac fermions. Recent work has extended this idea to realize a single unpaired Weyl fermion in a finite lattice. Here, we realize the same using a $2n$-dimensional string defect embedded in $(2n+2)$-dimensional spacetime on a finite lattice for $n=1$. This string is a lattice version of the continuum axion string described in Callan-Harvey \cite{Callan:1984sa}. Our results are obtained in a Hamiltonian formulation in Minkowski spacetime. Extending the results to Euclidean spacetime and to $n>1$ is straightforward. This work has applications to lattice chiral gauge theories and axion cosmology.  
\end{abstract}
\maketitle

\section{Introduction}
Formulating quantum field theories on the lattice is essential to extracting non-perturbative physics out of them. However, ensuring that the lattice path integral has the correct continuum limit is a nontrivial task. Among the many challenges that one faces while attempting to construct these lattice formulations, one pertains to formulating light chiral fermions. While lattice chiral fermions are primarily pursued for simulations in lattice gauge theory, their significance extends far beyond this field. In particular, chiral fermions play a crucial role in cosmology and early universe physics—appearing, for example, on topological defects such as cosmic strings and domain walls \cite{Callan:1984sa}, where they can significantly influence phenomenology \cite{ Copeland:1987th, Agrawal:2020euj, Fukuda:2020kym, Ibe:2021ctf}.
To date, the phenomenological implications of chiral fermions have been explored mostly in the continuum. However, lattice-regulated chiral fermions can alter these implications in substantial ways, especially since the lattice regulator typically modifies anomaly inflow in a nontrivial manner. For this reason, lattice chiral fermions are of broad theoretical and phenomenological interest.

It turns out formulating lattice chiral fermions with good chiral symmetry, whether this is a global symmetry or a gauged one,  is challenging. The obstacle can be traced back to the fermion doubling problem and the Nielsen-Ninomiya theorem \cite{Nielsen:1981xu, Nielsen:1980rz, Nielsen:1981hk}, which appears to preclude the existence of a single Weyl fermion or even a single massless Dirac fermion on the lattice without sacrificing some of the other desirable features of the theory, e.g. locality or chiral symmetry.
The challenge of realizing a massless lattice Dirac fermion with good global chiral symmetry was overcome using
the idea of domain wall fermions \cite{Kaplan:1992bt, Shamir:1993zy, Furman:1994ky} and overlap fermions in the nineties \cite{Neuberger:1997bg, Neuberger:1997fp, Narayanan:1993ss, Narayanan:1993sk}. The two approaches are in fact related to each other. For reasons that we will explain below, 
we will refer to these ideas as the extra dimension approach. 

The topic of lattice regulated chiral gauge theories, e.g. a lattice formulation of the standard model of particle physics, remains a topic of ongoing research. A consistent proposal for abelian lattice chiral gauge theories  was formulated by Luscher in \cite{Luscher:1998du}. However, its consistency for non-abelian theories has not been verified. Some of the other promising directions include symmetric mass generation,  \cite{Eichten:1985ft, Wen:2013ppa, You:2014sqa, You:2014vea, Wang:2018ugf, Catterall:2020fep, Wang:2022ucy, Razamat:2020kyf}, bosonization-based approaches \cite{Berkowitz:2023pnz}, gauge-fixed approaches \cite{Shamir:1995gd, Golterman:1996ph, Bock:1997fu, Bock:1997vc, Golterman:2004qv, Golterman:2005ha}, etc. 
In recent work, a lattice regulator for chiral gauge theories was proposed that relied on the extra dimension approach \cite{Kaplan:2023pxd, Kaplan:2023pvd} along the lines of domain wall fermions \cite{Kaplan:1992bt}. This paper concerns itself with generalizing this extra dimension idea beyond domain wall fermions. 

The conventional extra dimension approach to engineering a $2n$-dimensional massless lattice Dirac fermion is based on the idea of domain wall fermions (DWF). It involves introducing a massive Dirac fermion in $2n+1$ spacetime dimensions where one of the spatial dimensions, e.g. $x^{2n+1}$ is considered to be an extra dimension. The mass of this Dirac fermion is made to depend on the extra dimension $x^{2n+1}$ such that it changes sign at some $x^{2n+1}$, e.g. $x^{2n+1}=0$. It is well known  that the low energy spectrum of this theory realizes a $2n$-dimensional Weyl fermion  localized on the defect/domain wall at $x^{2n+1}=0$ \cite{Callan:1984sa, PhysRevD.13.3398, Kaplan:1992bt}. In a finite volume analysis with periodic boundary conditions, a wall necessarily appears with an anti-wall, which realizes an opposite chirality Weyl fermion. As a result the low energy spectrum of the two defects together is massless and Dirac-like. A similar spectrum can be obtained on the lattice with analogous mass defects for $(2n+1)$-dimensional Wilson fermions, as shown in \cite{Kaplan:1992bt}. This construction enables the realization of an arbitrary number of $(2n)$-dimensional massless Dirac fermions on the lattice by introducing as many $(2n+1)$-dimensional massive Dirac fermions with domain wall mass defects. This idea has been extensively used for simulating vector-like gauge theories, where the gauge interactions with right and left chirality fermions are identical,  e.g. quantum chromodynamics (QCD). This construction, however, is not suitable for chiral gauge theories where chiral symmetry is gauged, or in other words when the gauge field interaction with  fermions discriminates between different chiralities. However, recent work has demonstrated that the domain wall based extra dimension approach can be suitably modified to realize a single unpaired Weyl fermion on a finite lattice \cite{ Kaplan:2023pvd} which in turn can be used to formulate lattice chiral gauge theories \cite{Kaplan:2023pxd}. Index theorems and anomalies of this construction have been explored in \cite{Kan:2025njj, Aoki:2022aez}.

Although the most widely known higher-dimensional framework, as described above, uses a $(2n+1)$-dimensional spacetime manifold to engineer $2n$-dimensional chiral fermions, it is possible to achieve the same using a $(2n+2)$-dimensional manifold. In the latter case, the chiral fermions are localized on a defect known as an axion string \cite{Callan:1984sa}. The idea is to begin with a $(2n+2)$-dimensional fermion which allows one to write two types of ``mass" terms. One of them is the conventional mass term which we can refer to as a scalar mass term; the other is a pseudo-scalar mass term. When the two masses are uniform, one can redefine the fermion fields to set the pseudoscalar mass or the Dirac mass to zero while keeping the spectrum unchanged. However, this is not possible when the masses have spatially varying profiles. Interestingly, the two mass terms can be combined to mimic an axion-fermion coupling \cite{Callan:1984sa} by replacing the two masses by two real scalar/pseudo-scalar field operators. In the presence of an axion string where the ``axion" field winds by $2\pi$ around a $2n$-dimensional string defect embedded in $(2n+2)$-dimensional infinite spacetime, the low energy spectra again realizes a single Weyl fermion localized on the defect \cite{Callan:1984sa}. On the lattice the string defect can most easily be realized using crossed domain walls \cite{Sen:2022dkl, Sen:2023rrx, Kaplan:2021ewi, Kaplan:2022uoo}, where the scalar mass is $x^{2n+1}$-dependent and has a domain wall at $x^{2n+1}=0$ and the pseudo-scalar mass is $x^{2n+2}$-dependent with a domain wall at $x^{2n+2}=0$. The lattice analysis, of course, requires Wilson-like mass terms for the fermions, as discussed in \cite{Sen:2022dkl}. It was shown in \cite{Sen:2022dkl} that the low-lying spectra on the infinite spacetime lattice string matches that of the domain wall fermion on an infinite $(2n+1)$-dimensional lattice qualitatively, in the sense that both realize a single Weyl fermion at long wavelength. However, in a finite volume with periodic boundary conditions for the string, each domain wall will be accompanied by an anti-wall, and in general will result in two string and two anti-string configurations. The two strings will host two positive (right) chirality Weyl fermions and the two anti-strings will host two negative (left) chirality Weyl fermions. As a result, the low-lying spectra will diverge from that of the DWF in finite volume, realizing two massless Dirac fermions as opposed to a single massless Dirac fermion in the case of a DWF. The absence of a single massless Dirac fermion for the simple string defect on a finite space appears to preclude the realizability of an unpaired Weyl fermion on a finite lattice string. The goal of this paper is to circumvent these obstructions. We achieve this in two steps: 
\begin{enumerate}
\item We first demonstrate how one can engineer a single massless Dirac fermion using string defects in a finite lattice. 
\item We then show how this lattice defect can be modified to engineer a single Weyl fermion. 
\end{enumerate}
Although the entire analysis of this paper can be performed in Euclidean spacetime, we will work in Minkowski spacetime and Hamiltonian framework. We will denote $x^1$ as the time direction; the rest of the directions are spatial. 

The organization of the paper is as follows. We begin with a brief review of chiral fermions on domain walls and string defects in continuous infinite spacetime. We then discuss the same in an infinite spatial lattice. Since we are working in Minkowski Hamiltonian framework, we do not discretize time. This discussion is followed by a spatial lattice with finite volume for both the domain wall and the string defect. We contrast the low energy spectra of the two and highlight their differences. In the following section, we introduce modifications of the string defect to engineer a single massless Dirac fermion on a finite lattice string. This defect is then further modified to engineer a single Weyl fermion. We conclude with a discussion of our results.

\section{Chiral edge states in infinite space: continuum and lattice}
We first consider a
$(2n+1)$-dimensional massive Dirac fermion in infinite continuous space-time where the Dirac mass $m$ has the spatial profile
\beq
m=m_0\epsilon(x^{2n+1}).
\eeq
Here $\epsilon(x^{2n+1})$ gives the sign of $x^{2n+1}$. It is well known that the low energy spectrum of such a mass defect exhibits a Weyl fermion localized on the wall at $x^{2n+1}=0$ \cite{Callan:1984sa}. To achieve the same with a string defect in infinite $2n+2$ continuous dimensions, one can begin with a massive fermion $\psi$ in $2n+2$ dimensions with the action
\beq
    S = \sum_{\mu=1,\cdots 2n+2}\int d^{2n+2}x\,\,\bar{\psi}(i\Gamma^{\mu}\partial_\mu -(\phi_1-i\phi_2\bar{\Gamma}))\psi.
    \label{cont}
\eeq
Here $\Gamma^{\mu}$ are gamma matrices and $\bar{\Gamma}=i\Gamma^1\Gamma^2\cdots\Gamma^{2n+2}$. In $2n+2$ dimensions we can write two types of ``mass" terms:  the ``scalar" mass term, which we denote as $\phi_1$, and the ``pseudo-scalar mass" term, which we denote as $\phi_2$.  They can be combined into a complex mass $\phi=\phi_1+i\phi_2$. It was shown in \cite{Callan:1984sa} that when $\log\phi$ 
winds around a string defect localized at some  $x^{2n+1}, x^{2n+2}$, e.g. $x^{2n+1}=x^{2n+2}=0$, the low energy spectrum exhibits a Weyl fermion of a specific chirality localized on the string defect. Specifically, \cite{Callan:1984sa} considered an ansatz of the form $\phi= v(r) e^{i\theta}$ in cylindrical coordinates, where $\theta$ is the azimuthal angle $\arctan(x^{2n+2}/x^{2n+1})$ and $r$ is the radial coordinate with $r=\sqrt{(x^{2n+1})^2+(x^{2n+2})^2}$, and found a single Weyl fermion in the low energy spectrum. The details of the spectrum may change depending on the specific form of the defect. However, the Weyl nature of the spectrum is topological and persists as long as the phase of $\phi$ completes a $2\pi$ winding around the defect.
\begin{figure}[t]
    \centering
    \includegraphics[width=0.7\linewidth]{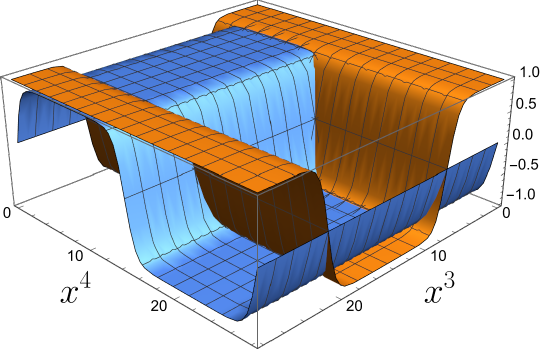}\vspace{.5 cm}
    \includegraphics[width=0.7\linewidth]{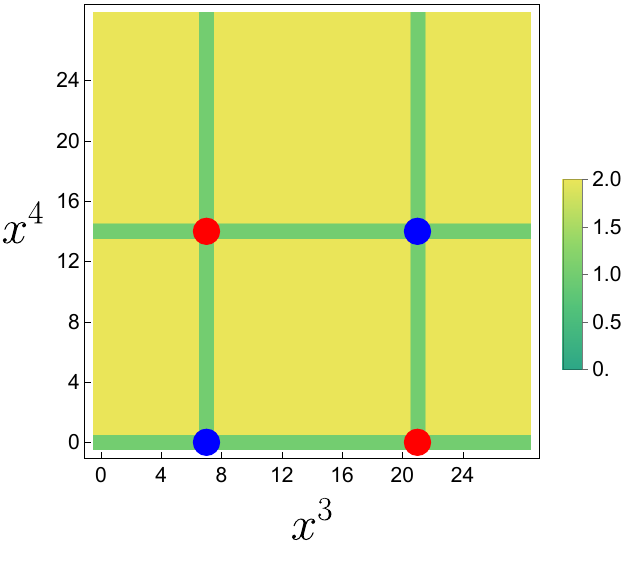}
    \caption{Top: Crossed domain walls (Eq. \ref{stringfin2}) producing two vortex-antivortex pairs for $N = 28$. $\phi_1$ is shown in  orange and $\phi_2$ is shown in blue. We have used a $\tanh$ profile for the domain walls instead of step functions. Bottom: A top-down view of $\phi_1^2 + \phi_2^2$, showing the domain wall defects and where they cross, i.e. the vortex defects. The vortices are colored blue, the anti-vortices red.}
    \label{fig:fourvortexfigure}
\end{figure}
\begin{figure}[t]
    \centering
    \includegraphics[width=0.7\linewidth]{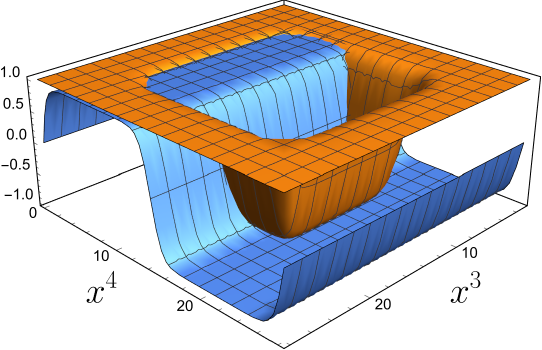}\vspace{.5cm}
    
    \includegraphics[width=0.7\linewidth]{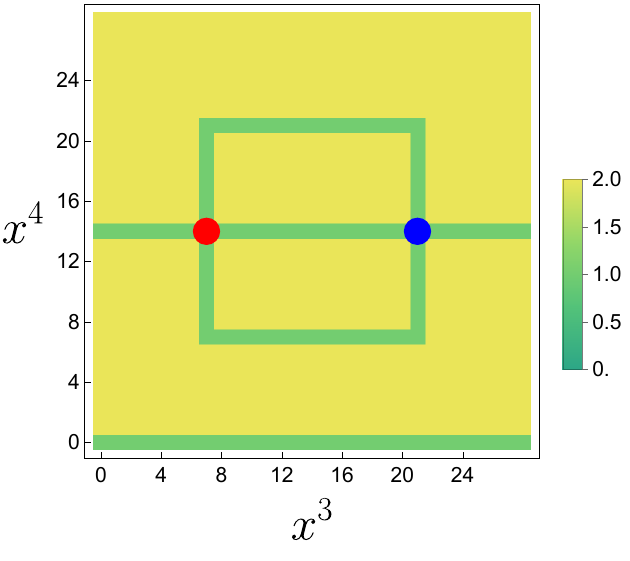}
    
    \caption{Top: The domain walls given by Eq. \ref{stringfin3}, \ref{eq:flatprofile} with $P$ given in Eq. \ref{4vor}, for $N = 28$. $\phi_1$ is shown in orange and $\phi_2$ is shown in blue. We have used $\tanh$ profile instead of step function for the domain wall illustration. Bottom: A top-down view of $\phi_1^2 + \phi_2^2$. The vortices are located at the blue (vortex) and red (anti-vortex) points, where the domain walls cross.}
    \label{fig:FlatWall}
\end{figure}
\begin{figure}[t]
    \centering
    \includegraphics[width=1\linewidth]{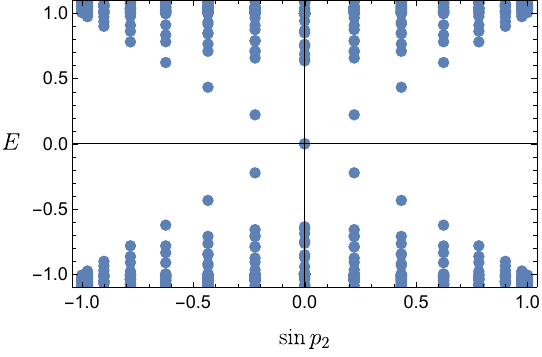}
    \caption{The low-lying energies for the string defects described by Eq. \ref{stringfin3}, \ref{eq:flatprofile} with $P$ given in Eq. \ref{4vor} for $N = 28$. The plot shown uses $p_2 = \frac{k\pi}{N}$ with $k$ going from $-N$ to $N-1$ for a lattice with $N = 28$.}
    \label{fig:2Ddisp}
\end{figure}
\begin{figure*}
    \centering
    \includegraphics[width=.7\linewidth]{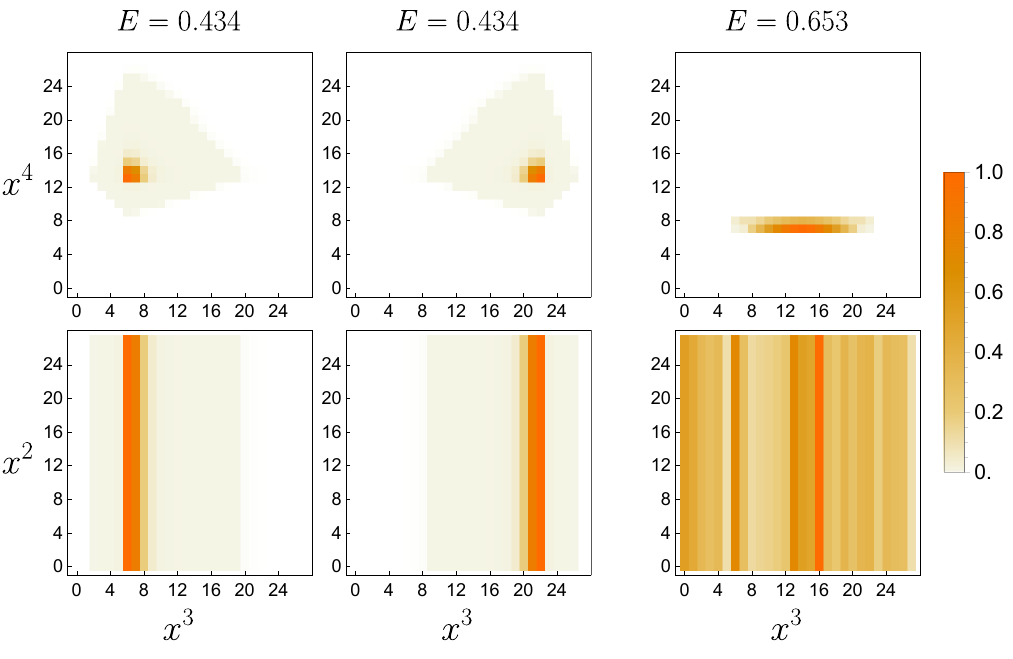}
    \caption{Fermion number charge density $\psi^\dagger \psi$ for some of the low-lying eigen states $\psi$ shown in Figure \ref{fig:2Ddisp}, with $N = 28$. We have normalized the fermion number charge density with respect to its maximum value in each plot. Bright orange indicates larger density. The top three plots are plotted in an $x^3x^4$ plane, and the bottom three plots are plotted in the $x^3 x^2$ plane for $x^4=N/2 = 14$. The mode in the left figures lies on the positive-chirality branch of the dispersion, while the mode in the center figures lies on the negative-chirality branch. In the right figures we are showing a mode which is not localized on the strings and has energy close to the cutoff.}
    \label{fig:2Ddensity}
\end{figure*}

{\bf Infinite lattice:} It was first pointed out in \cite{Kaplan:1992bt} that domain wall defects embedded in  $2n+1$ dimensions can be used to simulate lattice chiral fermions. The higher-dimensional theory in this case consists of Wilson fermion with a mass defect. Without loss of generality we can consider $n=1$. In infinite $2n+1=3$ dimensions with a two-dimensional spatial lattice the Wilson-Dirac Hamiltonian has the form 
\beq
    H_{\text{Wilson}}=\gamma^1\left(i\gamma^2\nabla_2+i\gamma^3\nabla_3+m +\frac{R}{2}(\Delta_2+\Delta_3)\right)
\eeq
where $\gamma^{1,2,3}$ are gamma matrices, the lattice spacing has been set to $1$ and for $i=2, 3$
\beq
    (\nabla_{i})_{x,x'}=\frac{\delta(x^{i}, (x^{'})^i-1)-\delta(x^{i}, (x^{'})^i+1)}{2},
\eeq
and
\beq
    \Delta_{i}=\delta(x^{i}, (x^{'})^i-1)+\delta(x^{i}, (x^{'})^i+1)-2\delta(x^{i}, (x^{'})^i).\nonumber\\
\eeq
$R$ is typically set to $1$.
We can make the choice $\gamma^1=\sigma_2$, $\gamma^2=i\sigma_1$, $\gamma^3=-i\sigma_3$ where $\sigma_i$ are Pauli matrices. The low energy spectrum includes localized chiral edge states \cite{Kaplan:2009yg, Jansen:1992tw, Jansen:1992yj}. We define $\sigma_3$ to be the chirality matrix. 
The chiral edge states have the form
\beq
&&\psi (x^2, x^3)\nonumber\\
&=&\sum_{p_2}\varphi(p_2) e^{i p_2 x^2} \begin{pmatrix}
(1-m_0\epsilon(x^3)+(1-\cos p_2))^{x^3}\\
0
\end{pmatrix}\nonumber\\
\eeq
These states are normalizable within the range $(1-\cos p_2)<m_0<2+(1-\cos p_2)$. If we set $m_0=1$, the defect realizes a single $(1+1)$-dimensional positive chirality Weyl fermion, i.e. eigenstates of the chirality operator $\sigma_3$ with an eigenvalue of $+1$ \cite{Kaplan:2009yg}. Its dispersion is given by $E=\sin p_2$ for $|p_2|<\pi/2$. 
\begin{figure}[t!]
    \centering
    \includegraphics[width=.9\linewidth]{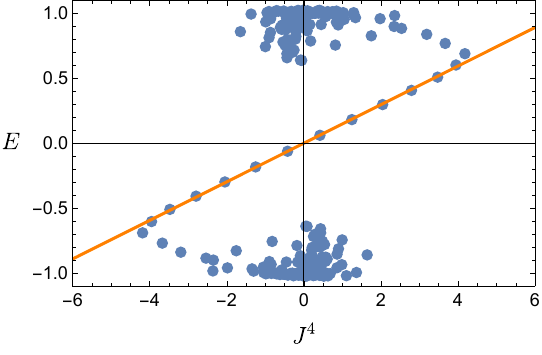}
    \caption{The dispersion for low-lying fermion modes with the crossed domain walls given by Eq. \ref{eq:curvedprofile}. We see a Weyl branch below $E \approx 0.6$. The states on this line are localized on the string defect.}
    \label{fig:JDisp}
\end{figure}

The analysis of a $2n$-dimensional defect embedded in $2n+2$ dimensions with an infinite lattice was worked out in \cite{Sen:2022dkl}. We specialize to $n=1$. Here, one constructs the string defect by introducing domain walls in $\phi_1$ and $\phi_2$ of the form 
\begin{equation}
    \begin{split}
        \phi_1=m_1\epsilon(x^3) \\
        \phi_2=m_2\epsilon(x^4).
    \end{split}
    \label{cross}
\end{equation}
It was shown in \cite{Sen:2022dkl} that the crossed domain wall defects specified by Eq. \ref{cross} create a vortex at $x^3=x^4=0$ such that the phase of $\phi$ winds by $2\pi$ around this defect. To see this, one can re-parametrize 
$\phi$ as 
\beq
    \phi=h(\cos\alpha + i \sin \alpha); h\equiv |\phi| 
\eeq
and show that $\int_\mathcal{C} \alpha \,\,dl=2\pi$ where $\mathcal{C}$ is a closed contour encircling the defect at $x^3=x^4=0$. The lattice analog of the action of Eq. \ref{cont} will include a fermion with Wilson terms. 
E.g. \cite{Sen:2022dkl} introduced Wilson-like terms, which result in the lattice Hamiltonian, 
\begin{equation}
    H_{\text{str}}=\Gamma^1(-i\Gamma^{2}\nabla_2-i\Gamma^{3}\nabla_3-i\Gamma^{4}\nabla_4+(\phi_1-i\phi_2\bar{\Gamma})+W)
\label{vorh}
\end{equation}
where $W$ has a form inspired by the standard Wilson term and is given by
\beq
    W=\frac{R}{2}\left(\Delta_2+\Delta_3-i\bar{\Gamma}(\Delta_2+\Delta_4)\right)
\eeq
with $R=1$. We will use the chiral basis for the gamma matrices 
\beq
    \Gamma^1=\begin{pmatrix}
        0 && 1\\
        1 && 0
    \end{pmatrix}, \quad
    \Gamma^{i+1}=\begin{pmatrix}
        0 && \sigma_i\\
        -\sigma_i && 0
    \end{pmatrix} ,
\eeq
for $i=1,2,3$. To measure the chirality of string modes, we use $\Gamma^{\text{str}} = \Gamma^1 \Gamma^2$.

As shown in \cite{Sen:2022dkl}, if we set $m_1=m_2=1$, the low energy spectrum in the background of this mass defect exhibits a Weyl fermion localized at $x^3=x^4=0$. Thus, we see that the spectrum of the domain wall defect embedded in $2n+1=3$-dimensional infinite spacetime and the string defect embedded in $(2n+2)$-dimensional infinite spacetime are qualitatively the same, in the continuum and on the lattice. 
\begin{figure*}
    \centering
    \includegraphics[width=.7\linewidth]{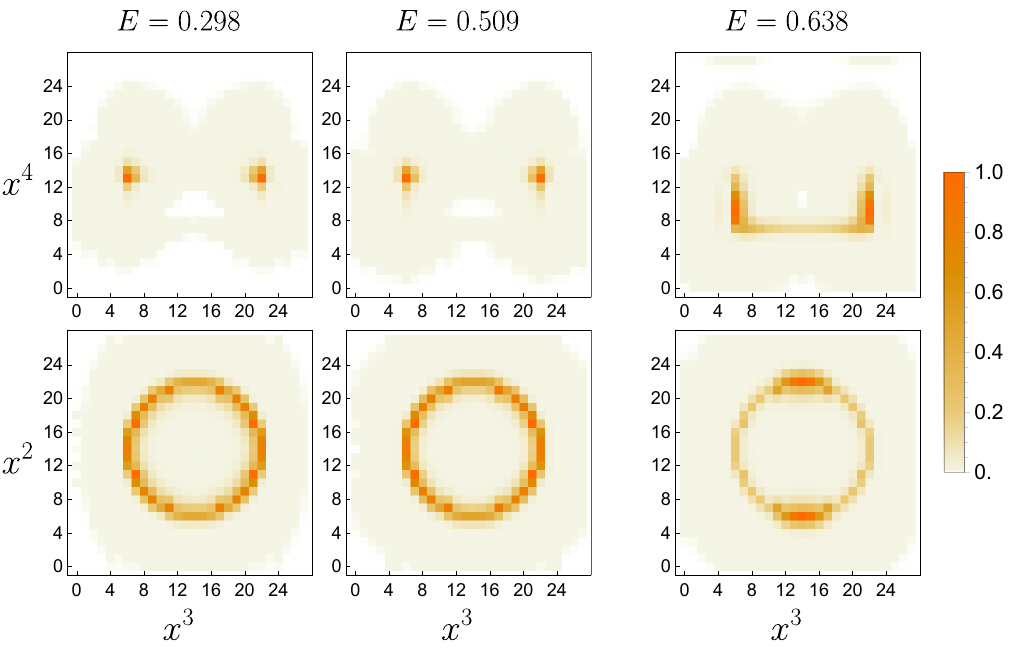}
    \caption{The fermion number charge density $\psi^\dagger \psi$ for some of the low-lying states $\psi$ shown in Fig. \ref{fig:JDisp}, with $N = 28$. We have normalized the fermion number charge density with respect to its maximum value in each plot. Bright orange indicates larger density. The top three plots are plotted in the $x^2 = N/2 = 14$ plane, and the bottom three plots are plotted in the $x^4 = N/2 = 14$ plane. The left and center figures are modes which lie in the linear Weyl region of the dispersion, and are well-localized on the singly connected vortex loop defect, cf. Fig. \ref{fig:2Ddensity}. The right figures show the lowest-energy mode not on the linear Weyl branch. It is not localized on the vortex loop.}
    \label{fig:3Ddensities}
\end{figure*}
\section{Chiral edge states on a Finite lattice} Let us now consider a finite spatial lattice with periodic boundary conditions (PBC) for all spatial directions. The lattice size in each spatial direction can be taken to equal $L$, identifying $L/2$ with $-L/2$. If we attempt to embed a domain wall defect in the extra dimension $x^3$ for $n=1$, we inevitably end up with a wall and an anti-wall due to the periodic nature of the manifold, e.g.
\beq
    m(x^3)=m_0\epsilon(x^3)-2m_0\Theta(x^3-L/2) ,
\label{mdw}
\eeq
where $\Theta$ is a step function taking the value of $+1$ for positive argument and 0 for negative argument. Thus we are left with a wall at $x^3=0$ and an antiwall at $x^3=\pm L/2$, the separation between the two defects being $L/2$.
As a result, with $m_0=R=1$ we realize a positive (right) chirality Weyl fermion on the wall as well as a negative (left) chirality Weyl fermion on the anti-wall. The wall and the anti-wall together lead to the realization of a single massless Dirac fermion in the low energy spectrum.

Placing the string defect of Eq. \ref{cross} on a periodic lattice will, however, lead to a spectrum that is qualitatively different from the above domain wall case. In the case of the string, each of the domain walls, one for $\phi_1$ and the other for $\phi_2$, will lead to an anti-wall resulting in four string defects, 
e.g.
\begin{equation}
    \begin{split}
        \phi_1=m_1\epsilon(x^3)-2m_1\Theta(x^3-L/2) \\
        \phi_2=m_2\epsilon(x^4)-2m_2\Theta(x^4-L/2) .
    \end{split}
    \label{stringfin}
\end{equation}

Two of the strings have positive net winding for the phase of $\phi$ and are located at $x^3=x^4=0$ and $x^3=x^4=L/2$: we can refer to these as vortices. The other two are located at $x^3=0, x^4=L/2$ and $x^3=L/2, x^4=0$, and will have net negative winding: we can refer to these as anti-vortices. As a result, the low energy spectrum will include two Weyl fermions of each chirality: positive (right) ones located on the vortices and negative (left) ones on the anti-vortices. Thus, the spectrum realizes two massless Dirac fermions as opposed to a single massless Dirac fermion.  

Finally, the program of DWF, i.e. engineering a massless  $2n$-dimensional Dirac fermion by embedding mass defects in a $(2n+1)$-dimensional spacetime 
manifold was modified to realize a single unpaired Weyl fermion on a finite lattice in \cite{Kaplan:2023pvd}. For $n=1$, the modification can be implemented by a mass defect of the form 
\beq
    m(x^2, x^3)=\begin{cases} 
        m_0 &\sqrt{(x^2)^2+(x^3)^2}\leq r_0\\
        -m_0 &\sqrt{(x^2)^2+(x^3)^2}> r_0\\
    \end{cases} .
    \label{circ}
\eeq
This engineers a disk defect with a circular boundary and can be placed in a manifold with PBC in $x^2$ and $x^3$ without encountering another defect, as shown in \cite{Kaplan:2023pxd, Kaplan:2023pvd}. For $m_0=1$ with $R=1$ the circular defect realizes a single Weyl fermion localized at $r=\sqrt{(x^2)^2+(x^3)^2}=r_0$ that circulates in the azimuthal direction. One could rephrase the idea from the point of view of conventional domain wall as follows. Consider the defect of Eq. \ref{mdw} and restrict it to hold only for $x^2=0$, i.e.  
\beq
m(0, x^3)=m_0\epsilon(x^3)-2m_0\theta(x^3-L/2)
\eeq
Then the separation between the two defects is $L/2$ at $x^2=0$. One can now demand that as we move away from the slice of spacetime at $x^2=0$, we retain the two mass defects in $x^3$, but reduce their $x^3$ separation with increasing $|x^2|$. Eventually the separation reduces to zero, which implements a singly connected boundary/defect instead of two disconnected boundaries as in Eq. \ref{mdw}. This is what is implemented by Eq. \ref{circ}. The difference of topology between the two cases, i.e.  two disconnected defects/boundaries in Eq. \ref{mdw} vs a singly connected defect/boundary in Eq. \ref{circ}, leads to a crucial difference in the spectra, namely a pair of opposite chirality Weyl fermions, i.e. a Dirac fermion in the former vs a single unpaired Weyl fermion in the latter.

If we apply an analogous modification to the string defect of \ref{stringfin}, e.g. by demanding that the separation between the vortex string at $x^3=x^4=0$ and the one at $x^3=0, x^4=L/2$ reduces as a function of $|x^2|$, while being maximum at $x^2=0$ and disappearing for some finite $|x^2|$, and similarly, the separation between the vortex string at $x^3=L/2, x^4=0$ and the one at $x^3=L/2, x^4=L/2$ reduces as a function of $|x^2|$, while being maximum at $x^2=0$ and disappearing for some finite $|x^2|$, we end up with two singly connected defects. There appears to be no way to reduce the number of singly-connected defects down to $1$. The absence of a single unpaired singly connected defect means that without further modification, the defect of Eq. \ref{stringfin} cannot realize a single unpaired Weyl fermion.

This leads us to the goal of this paper: to modify the defects of Eq. \ref{cross} and \ref{stringfin} in such a way as to match onto the spectrum of a single massless Dirac fermion, similar to that produced by the defect in Eq. \ref{mdw}, and then further modify the defect along the lines of \cite{Kaplan:2023pvd} to engineer a single unpaired Weyl fermion.

\section{New defect}
We first construct a defect that realizes a single massless Dirac fermion at low energy on a manifold with PBC. We will then modify this defect to engineer an unpaired Weyl fermion spectrum.
\subsection{Dirac fermion}
Before we write the profile for the modified defect, for convenience we will rewrite the defect in Eq. \ref{stringfin} in a different form. We begin by shifting 
\beq
    x^3\rightarrow x^3-L/4
\eeq
and set $m_1=m_2=1$. Then the defect of Eq. \ref{stringfin} becomes 
\begin{equation}
    \begin{split}
        \phi_1 &= \epsilon(x^3-L/4)-2\Theta(x^3-3L/4) \\
        \phi_2 &= \epsilon(x^4)-2\Theta(x^4-L/2) .
    \end{split}
    \label{stringfin2}
\end{equation}

Anticipating the form of the modification needed to engineer a single massless Dirac fermion, we choose to re-express the  profile for $\phi_1$ in Eq. \ref{stringfin2} in a slightly altered form while keeping the form of $\phi_2$ unchanged:
\begin{equation}
    \begin{split}
        \phi_1(x^3,x^4) &= 1 + h(x^4) P(x^2, x^3), \\
        \phi_2(x_4) &= \epsilon(x^4)-2\Theta(x^4-L/2) 
        \end{split}
    \label{stringfin3}
\end{equation}
where we set $P(x^2,x^3)$ and $h(x^4)$ to
\begin{equation}
    \begin{split}
        P(x^2,x^3) &= \Theta\left(1-\frac{(x^3-L/2)^2}{(L/4)^2}\right), \\
        h(x^4) &= -2.
    \end{split}
    \label{4vor}
\end{equation}

We provide a figure showing the mass profile for $\phi_1$ and $\phi_2$ in Fig. \ref{fig:fourvortexfigure}. To produce the figure, we consider a spatial square lattice of lattice spacing $1$. The number of lattice sites in each direction is $N$. Therefore, the lattice profile for $\phi_1$ and $\phi_2$ is simply given by Eq. \ref{stringfin3}, \ref{4vor} with $L$ replaced by $N$. We take $N=28$ and use $\tanh$ in place of step functions for illustration. The profile of Eq. \ref{stringfin3}, \ref{4vor} includes four string defects, i.e. lines where the two domain walls cross. Two of the string defects are vortices with positive winding (colored blue) and two are anti-vortices with negative winding (colored red). For the purpose of illustration, in the figure we have replaced the step functions in the mass profiles with a smoother $\tanh$ function.

We now introduce a modified mass profile by altering $h$ to 
\begin{equation}
    h(x^4) = (\epsilon(x^4 - 3L/4) - \epsilon(x^4 - L/4))
    \label{eq:flatprofile}
\end{equation}
in Eq. \ref{stringfin3} without altering the form of $P$.
We provide a corresponding picture in Fig. \ref{fig:FlatWall}. It shows the mass profile on the top panel and the locations of the defects in the bottom panel. Once again, we have used $\tanh$ in place of step functions. This defect retains a single vortex and a single anti-vortex separated in $x^3$ by a distance of $L/2$. We color the vortex blue and the anti-vortex red in Fig. \ref{fig:FlatWall}. We expect to find a Weyl fermion of positive (right) chirality on the vortex and a Weyl fermion of negative (left) chirality on the anti-vortex, resulting in a single Dirac fermion in the low energy spectrum. 

To verify this explicitly, we will have to diagonalize the Hamiltonian $H_{\text{str}}$ in Eq. \ref{vorh} to obtain the spectrum. Fourier transforming in the $x^2$ direction we can write the Hamiltonian as

\begin{equation}
    \begin{split}
        H_{\text{str}} = \Gamma^1(&\Gamma^{2}\sin(p_2)-i\Gamma^{3}\nabla_3-i\Gamma^{4}\nabla_4 \\
        &+ (\phi_1-i\phi_2\bar{\Gamma})+W)
    \end{split}
    \label{vorh2}
\end{equation}
where 
\beq
    W=\frac{R}{2}\left(\Delta_3-i\Delta_4\,\,\bar{\Gamma}\right)
    +R(\cos p_2-1)(1-i \bar\Gamma) .
\eeq

We then diagonalize $H_{\text{str}}$ on an $N^3$ lattice with $N=28$. The corresponding low energy spectrum is shown in Fig. \ref{fig:2Ddisp}. The momenta $p_2$ take values $p_2 = \frac{k\pi}{N}$, where $k$ takes integer values from $-N$ to $N-1$. As expected for a Dirac fermion on a finite lattice, every energy eigenvalue, except the lowest, is doubly degenerate. In the infinite volume/continuum limit the lowest energy eigenvalue approaches zero and is also doubly degenerate, as expected for a continuum Dirac fermion. We find the cutoff scale to be $E \sim 0.63$ in units of the inverse lattice spacing.

We expect the low lying modes to be localized on the vortex defects, whereas the modes at the cutoff and above have no reason to be so. To illustrate this, we compute the fermion number charge density $\psi^{\dagger}\psi$ in 
Fig. \ref{fig:2Ddensity} for some of the low lying eigenstates $\psi$. These states have energy well below the cutoff of $\sim 0.63$. In particular, we plot a normalized fermion number charge density, defined as the fermion number charge density divided by its maximum value. 
As seen from the figure, these low-lying states are well localized on the string defects. We have also shown a state which is not localized on any of the string defects, instead appearing to localize on one of the domain walls. It should be noted that this state has an energy close to the cutoff scale.

\subsection{Weyl fermion}
In the previous section we were able to engineer a single pair of string defects, a vortex and an anti-vortex that stretch in the $x^2$ direction. Thus translational invariance in $x^2$ was preserved. This allowed us to construct the dispersion plotted in Fig. \ref{fig:2Ddisp}.
In order to engineer a singly connected defect, as required for an unpaired Weyl fermion, we break translation invariance in $x^2$. More specifically, we demand that the separation between the two defects, the vortex and the anti-vortex, become a function of $x^2$, being maximum in the $x^2=L/2$ plane, reducing as a function of growing $|x^2-L/2|$, and then vanishing at some $|x^2-L/2|\neq 0$. To implement this we modify the expression of $P$ in Eq. \ref{4vor} to
\begin{equation}
    P(x^2, x^3) = \Theta\left(1-\frac{(x^3-\frac{L}{2})^2}{(L/4)^2} - \frac{(x^2-\frac{L}{2})^2}{(L/4)^2}\right) .
    \label{eq:curvedprofile}
\end{equation}
This profile mimics Eq. \ref{circ} for $r_0 = L/4$, and implements an $x^2$-dependent separation between the two strings (vortex and anti-vortex), their separation being $L/2$ at $x^2=L/2$, and in general given by 
\beq
    D=\sqrt{\left(2 x^2-\frac{L}{2}\right)\left(\frac{3L}{2}-2x^2\right)}
\eeq

This produces the desired unpaired singly connected defect we were aiming to construct. We can refer to it as a vortex loop.
We can now diagonalize the Hamiltonian $H_{\text{str}}$ given this mass profile. Since we have broken translation invariance in $x^2$, a dispersion relating energy eigenvalues $E$ with momentum  $p_2$ is not meaningful. Instead, following \cite{Kaplan:2023pvd}, we compute the expectation value of the angular momentum operator in the $x^4$ direction for the eigenstates of the Hamiltonian. The angular momentum operator $J^4$ is given by
\begin{equation}
    \begin{split}
        J^4 &= L^4 + \frac{1}{2} S^4 , \\
        L^4 &=-i( x^2 \nabla_3 - x^3 \nabla_2) , \\
        S^4 &= \frac{i}{2}[\Gamma^2,\Gamma^3] = \mathbb{I}_2 \otimes \sigma_3 .
    \end{split}
\end{equation}

\begin{figure}[t!]
    \centering
    \includegraphics[width=0.85\linewidth]{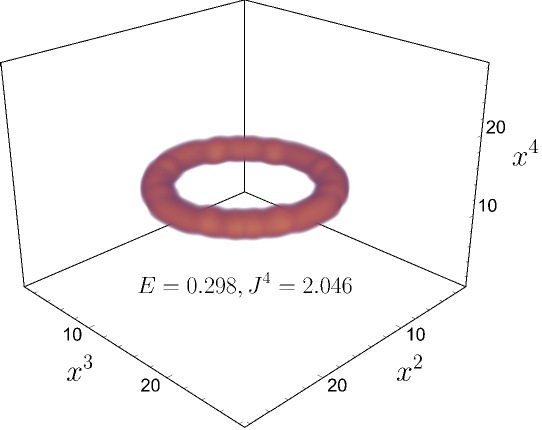}
    \caption{The fermion number charge density $\psi^\dagger \psi$ for one of the modes on the Weyl branch of Fig. \ref{fig:JDisp} with $N = 28$. 2D slices for $x^4 = 14$ and $x^2 = 14$ are shown in the left column of Fig. \ref{fig:3Ddensities}.}
    \label{fig:3DVortex}
\end{figure}

We use an $N^3$ lattice of size $N = 28$ with PBC in all of the spatial directions and plot the dispersion in Fig. \ref{fig:JDisp}. For $E \lesssim 0.6$ we have a right-handed Weyl-like dispersion relation. The associated states are localized on the string defects. This is illustrated in Fig. \ref{fig:3Ddensities}, where we have again computed the normalized fermion number charge density. The upper panel shows the density profile in the $x^2=N/2$ plane, whereas the lower panel shows the density in the $x^4=N/2$ plane for three different states, two of which lie on the Weyl branch with energy well below the cutoff; the remaining state has energy just above the cutoff. The former are well localized on the defect while the latter is not. One can construct an anti-vortex loop by simply flipping the sign of one of the masses $\phi_1$ or $\phi_2$. On the anti-vortex loop we find a negative chirality Weyl fermion. 

In Fig. \ref{fig:3DVortex} we plot the fermion number charge density for one of the low-lying Weyl modes to show its localization in three dimensional space. As seen from the figure, the state is localized along the core of the vortex.

\section{conclusion}
In this paper we demonstrated how one can engineer a $2n$-dimensional massless lattice Dirac fermion and a $2n$-dimensional unpaired lattice Weyl fermion by embedding a $2n$-dimensional mass defect  in $2n+2$ dimensions. These defects are analogs of the axion string defect in the continuum. The results of this paper suggest that, much like the domain wall mass defect in $2n+1$ dimensions, lattice axion strings embedded in $2n+2$ dimensions can potentially be used to regulate chiral gauge theories. To implement this, one has to begin with several $(2n+2)$-dimensional massive Dirac fermions with coinciding mass defects. More specifically, every Dirac fermion will be associated with either a vortex loop (a defect with positive winding) or an anti-vortex loop (negative winding). Thus, the coinciding vortex/anti-vortex loop defect will host either a right or a left chirality Weyl fermion for each of the $(2n+2)$-dimensional Dirac fermions. Then one can engineer the Weyl fermion content of the desired chiral gauge theory by introducing higher-dimensional Dirac fermions with appropriate vortex or anti-vortex defects. One can then gauge the appropriate anomaly-free chiral symmetry group of interest. Note that, for the purpose of application to chiral gauge theories, the gauge fields have to be $2n$-dimensional: they fluctuate on the defect with the $2n$-dimensional gauge field action and propagate only in $2n$ dimensions. Away from the loop defect, one will have to continue the gauge fields into the $(2n+2)$-dimensional bulk in a way that preserves $2n$-dimensional gauge invariance. A natural choice may be to use the $(2n+2)$-dimensional gauge field equation of motion to continue the defect gauge fields into the bulk. The details of adding the gauge field is left to future work.  
Furthermore, it will be interesting to explore the Euclidean spacetime behavior of this defect in the presence of gauge fields. In particular, we expect zero modes in the spectrum if the gauge field configuration on the defect has a net instanton number. 

Finally, axion strings/loops and the localized chiral edge states are important to the phenomenology of the early universe, e.g. that of superconducting cosmic strings \cite{ Copeland:1987th, Agrawal:2020euj, Fukuda:2020kym, Ibe:2021ctf}. Thus far, phenomenological studies have focused on chiral edge states of continuum axion strings. Lattice regularized axion strings may have a much richer phenomenology. In particular, it is known that varying the parameters $m_1, m_2$ with respect to the Wilson parameter can alter the number of localized Weyl fermions and the net chirality on lattice axion strings \cite{Sen:2022dkl}. Evidently, the associated anomaly inflow and the Goldstone-Wilczek current will be altered as well \cite{Sen:2022dkl}. This is likely to impact the properties of the superconducting cosmic strings
and is worth investigating.
\section{Acknowledgement}
S.S.\ acknowledges support from the U.S.\ Department of Energy, Nuclear Physics Quantum Horizons program through the Early Career Award DE-SC0021892. J.K.\ acknowledges support from the U.S.\ Department of Energy under Grant No. DE-SC0023692.

\bibliography{refs.bib}

\end{document}